\documentclass[conference]{IEEEtran}
\IEEEoverridecommandlockouts
\usepackage{cite}
\usepackage{amsmath,amssymb,amsfonts}
\usepackage{algorithmic}
\usepackage{graphicx}
\usepackage{float}
\usepackage{textcomp}
\usepackage{xcolor}
\def\BibTeX{{\rm B\kern-.05em{\sc i\kern-.025em b}\kern-.08em
    T\kern-.1667em\lower.7ex\hbox{E}\kern-.125emX}}
\begin{document}

\title{Comparative Study of Bitcoin Price Prediction \\
}

\author{\IEEEauthorblockN{1\textsuperscript{st} Ali Mohammadjafari \\ Ali.mohammadjafari1@louisiana.edu\\
         School of Computing and Informatics\\
        University of Louisiana at Lafayette,
        USA}
 \\

}

\maketitle

\begin{abstract}

Prediction of stock prices has been a crucial and challenging task, especially in the case of highly volatile digital currencies such as Bitcoin. This research examineS the potential of using neural network models, namely LSTMs and GRUs, to forecast Bitcoin's price movements. We employ five-fold cross-validation to enhance generalization and utilize L2 regularization to reduce overfitting and noise. Our study demonstrates that the GRUs models offer better accuracy than LSTMs model for predicting Bitcoin's price. Specifically, the GRU model has an MSE of 4.67, while the LSTM model has an MSE of 6.25 when compared to the actual prices in the test set data. This finding indicates that GRU models are better equipped to process sequential data with long-term dependencies, a characteristic of financial time series data such as Bitcoin prices. In summary, our results provide valuable insights into the potential of neural network models for accurate Bitcoin price prediction and emphasize the importance of employing appropriate regularization techniques to enhance model performance.

\end{abstract}


\section{Introduction}
\subsection{Motivation and Overview}

Bitcoin is a digital currency that operates through a decentralized online network and does not depend on a government or legal entity. Its security and authenticity are maintained using cryptography and peer-to-peer networking\cite{yu2022bitcoin}. 

\begin{figure}[h]
    \centering
    \includegraphics[width=.43\textwidth, height=0.3\textwidth]{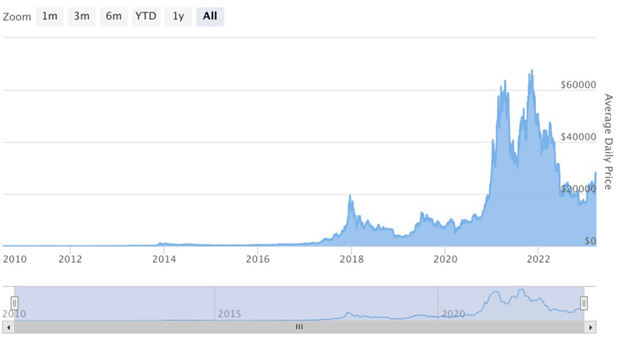}
    \caption{Bitcoin daily prices on Bitstamp (USD) from 29 November 2011 to 27 March 2023.
The upper line shows log prices, whereas the lower line shows plain prices. Some recurring patterns
seem to exist when considering the log value of the Bitcoin price.}
\label{fig:bitprice1}
\vspace{-.3 cm}
\end{figure}

At present, Bitcoin stands as the top-performing cryptocurrency across various categories such as Litecoin, Ripple, Bitcoin, and Ethereum. These cryptocurrencies utilize comparable algorithms for transactions and are known for their robust security. Furthermore, each cryptocurrency implements its unique set of rules and regulations to attract a wide range of investors at different times.\cite{kitamura2022can}. Figure~\ref{fig:bitprice1} shows the Bitcoin daily prices from 29 November 2011 to 27 March 2023 on Bitstamp (https://www.investing.com/), which is the longest-running cryptocurrency exchange.

Forecasting is a critical component of many financial sectors, as financial losses can cause significant problems for an industry's capital. There are numerous applications of forecasting, including weather prediction, stock market prediction, and natural disaster prediction such as floods and earthquakes.\cite{kumar2022efficient}.

Predicting the price of Bitcoin is essential for various reasons, particularly in helping investors and traders make informed decisions regarding the purchase or sale of Bitcoin. With the ability to forecast future price movements, investors can capitalize on profitable opportunities and minimize potential losses. Forecasting also enables businesses and individuals who accept Bitcoin as payment to plan their finances and budget accordingly.

The volatile nature of Bitcoin emphasizes the significance of forecasting. Since Bitcoin is not tied to any government or central bank, its value can change rapidly based on market demand and other factors. Precisely predicting these fluctuations can help investors and traders take advantage of these changes and avoid potential losses. 

Various researchers have examined several factors that affect the price of Bitcoin and analyzed the patterns of its fluctuations using diverse analytical and experimental techniques. Accurately forecasting Bitcoin prices is a challenging undertaking that necessitates advanced techniques such as machine learning and neural networks. Consequently, many prediction models for Bitcoin's price have been created based on deep learning techniques\cite{ren2022past}\cite{corbet2019cryptocurrencies}.

\subsection{Goal and Contributions}

This project aims to leverage two distinct neural network models to forecast Bitcoin prices and compare their effectiveness. To achieve this objective, the project will employ the Bitcoin Historical Data dataset from the Yahoo Finance website, which comprises daily Bitcoin prices ranging from 2015 to the present day. Additionally, the dataset contains other features such as volume and market capitalization that can be utilized as input variables to train neural network models.

The initial neural network model to be used in this project is a Long Short-Term Memory (LSTM) network. LSTMs are a type of recurrent neural network that are designed for time series prediction tasks. The LSTM model will be trained on a portion of the Bitcoin Historical Data dataset, employing past Bitcoin prices and additional features to anticipate future prices.

The other neural network model to be studied is Gated Recurrent Unit (GRU). GRU is a specialized design of recurrent neural network (RNN) that aims to address some of the limitations of conventional RNNs. One of the main drawbacks of RNNs is the "vanishing gradient problem," which occurs when gradients used to update the network weights during backpropagation become too small to effectively capture long-term dependencies. GRU will also be trained on the Bitcoin Historical Data dataset, leveraging a sliding window technique to anticipate future prices based on past prices and additional features.

To evaluate the performance of the two models, several metrics will be used, such as mean squared error (MSE) and mean absolute error (MAE). The models will also be compared using visualizations, such as line charts and scatter plots, to compare the predicted prices to the actual prices. We expecte that the result show that our models predict the future price of the bitcoin with 80 percent accuracy. 

In summery, the contributions of this work are as follows:
\begin{itemize}
    \item We develop two different neural network models (LSTM and GRU) to predict Bitcoin prices and compare their performance.
    \item We use the Bitcoin Historical Data dataset from Yahoo Finance, and use LSTM and GRU models to make predictions. 
    \item Based on the result, we provide insights into the effectiveness of these models for Bitcoin price prediction, which can have important implications for investors and traders in the cryptocurrency market.
    \end{itemize}
    The rest of this paper is organized as follows. Section II discusses background study and related prior works. We explain the overview of chosen neural network models architecture in Section III. Next, we discuss experimental evaluation and performance analysis in Section IV. Finally, Section V concludes the paper and provides a few avenues for the future studies.

\section{Related Works}
\subsection{Price prediction}
Price prediction using neural network models is growing in popularity due to the availability of more data and computing power. Neural networks, which are designed to resemble the human brain, can learn and recognize patterns in data. These models analyze historical price data to identify patterns and relationships that may not be immediately apparent to human analysts, and use this information to generate predictions about future price movements. However, such models require large amounts of historical price data and other relevant factors such as economic indicators, market sentiment, and news events. Hyper Parameters such as the number of layers, number of neurons per layer, and learning rate may also need to be carefully adjusted to achieve optimal performance\cite{chen2022china}.
 
 Cheng et al. have developed a method that uses a two-phase attention mechanism to jointly optimize inner-modality and inter-modality sources, allowing end-users to examine their importance. This approach provides investors with an interpretable and profitable option, enabling them to make informed investment decisions\cite{cheng2022financial}.
 
 Wu, Shengting, et al. incorporate investors' sentiment index, technical indicators, and stock historical transaction data as the feature set for stock price prediction using a long short-term memory network. The experiments show that the mean absolute error can achieve 2.386835, which is better than traditional methods\cite{wu2022s_i_lstm}.

So, we can see that By using neural networks to analyze large amounts of data, traders and analysts can gain valuable insights into market trends and make more informed trading decisions.

\subsection{Bitcoin Price prediction}
Aghashahi and Bamdad describe their approach to bitcoin price prediction, which involves using a combination of technical attributes such as price-related and lagged features as inputs for neural networks. By doing so, they aim to improve prediction accuracy and ultimately enhance profitability\cite{aghashahi2022analysis}. 

In order to improve the accuracy of Bitcoin price prediction, Ye, Zi, et al introduce a new ensemble deep learning model that uses both long short-term memory (LSTM) and gate recurrent unit (GRU) neural networks, along with technical indicators, price data, and sentiment indexes. The model is designed for near-real-time prediction of Bitcoin prices and performs better than daily predictions, with a mean absolute error (MAE) that is 88.74 percent lower\cite{ye2022stacking}.

Experimental results in the Ehsan Sadeghi, et al. work in 2022 demonstrated these findings that the 1-dimensional convolutional neural network and stacked gated recurrent unit model (1DCNN-GRU) showed superior performance compared to existing methods for Bitcoin, Ethereum, and Ripple datasets, with the lowest RMSE values of 43.933, 3.511, and 0.00128, respectively\cite{pour2022cryptocurrency}. 

In conclusion, various studies have been conducted on Bitcoin price prediction using neural network models. These studies have explored the use of various input features such as technical indicators, sentiment analysis, and lagged features to improve the prediction accuracy of neural network models. Additionally, the application of different types of neural networks such as long short-term memory (LSTM) and Gated Recurrent Unit (GRU) have been investigated to further enhance the predictive performance. Overall, the experimental results suggest that these models have shown promising results in predicting Bitcoin prices. However, further research is needed to improve the robustness and generalist of these models.

\section{Methodology}
\subsection{Problem Statement}

The popularity of cryptocurrency and the instability of the Bitcoin market have resulted in an increasing interest in predicting Bitcoin prices. This is because accurate predictions can guide investors to make informed decisions and minimize trading risks. However, forecasting the future value of Bitcoin is a complex task due to the market's highly intricate and nonlinear nature. Thus, this research aims to evaluate the performance of two different neural network models, namely LSTM and GRU, in predicting the future price of Bitcoin by leveraging past prices and other features. The study aims to compare the effectiveness of these models in terms of accuracy and provide valuable insights into their applicability to Bitcoin price prediction. The ultimate objective of the research is to support the development of more reliable and precise models for Bitcoin price forecasting, which can be beneficial for investors and researchers in the cryptocurrency domain.
\subsection{Initial Experiment}
Experimental Procedure:
\begin{itemize}
    
\item Data Preparation:
The Bitcoin Historical Data dataset will be downloaded from Yahoo Finance website. By preprocessing, data is ready to process by network models. By using 5-fold cross-validation, we can obtain more accurate and reliable estimates of the model's performance on unseen data 
\item Neural Network Models:
we train LSTM and GRU models on a given dataset and evaluate their performance on test data. Additionally, to generalization of the models, we apply L2 regularization to the weights of the models. To assess generalization of the models, we use 5-fold cross-validation. 

\item Performance Comparison:
The performance of the LSTM and GRU models will be compared using various metrics, such as mean squared error (MSE), mean absolute error (MAE), and accuracy. The models will be evaluated on the validation set and the testing set.
\item Conclusion:
Based on the results of the experiment, we will conclude which neural network model performs better for Bitcoin price prediction. The findings of this study can be useful for investors and traders who are interested in predicting the future price of Bitcoin.

\end{itemize}

\subsection{Data Processing}

In this section, the method of gathering Bitcoin price data and transforming it for use as training data and testing data is explained. The data is obtained from the Yahoo Finance website (https://finance.yahoo.com/quote/BTC-USD). As the data is originally combined with different information, it requires preprocessing. 

\subsubsection{ Data Collection and Preprocessing}
\begin{figure}[h]
    \centering
    \includegraphics[width=0.5\textwidth, height=0.2\textwidth]{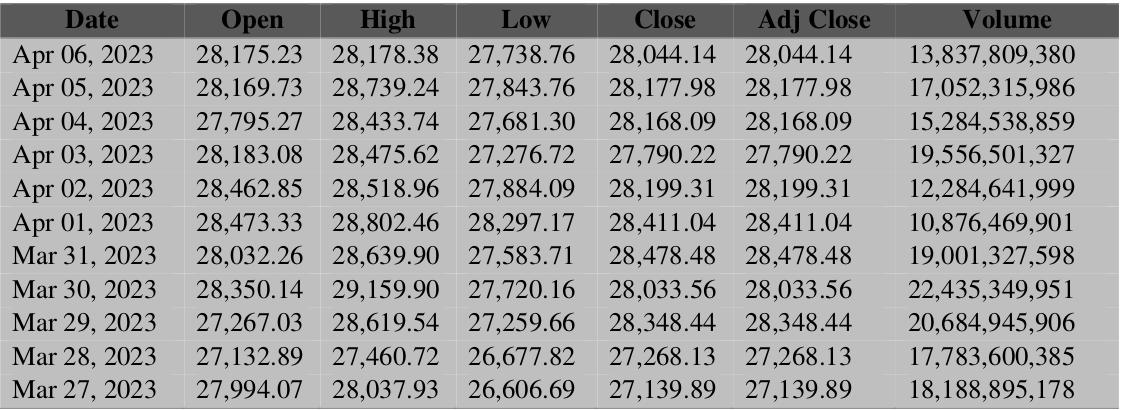}
    \caption{ Display of the Data collected}
\label{fig:data}
\vspace{-.3 cm}
\end{figure}

We collected Bitcoin price from Yahoo Finance. The collected data has six features: open
price, close price, high price, low price, adjust close price and volume at each time period (i.e., dailly). The price and volume data were collected from December 31, 2015 to April 6, 2023. Figure~\ref{fig:data} displays the few data collected for this research. 

\subsubsection{ K-Fold Cross Validation}
\begin{figure}[h]
    \centering
    \includegraphics[width=0.5\textwidth, height=0.2\textwidth]{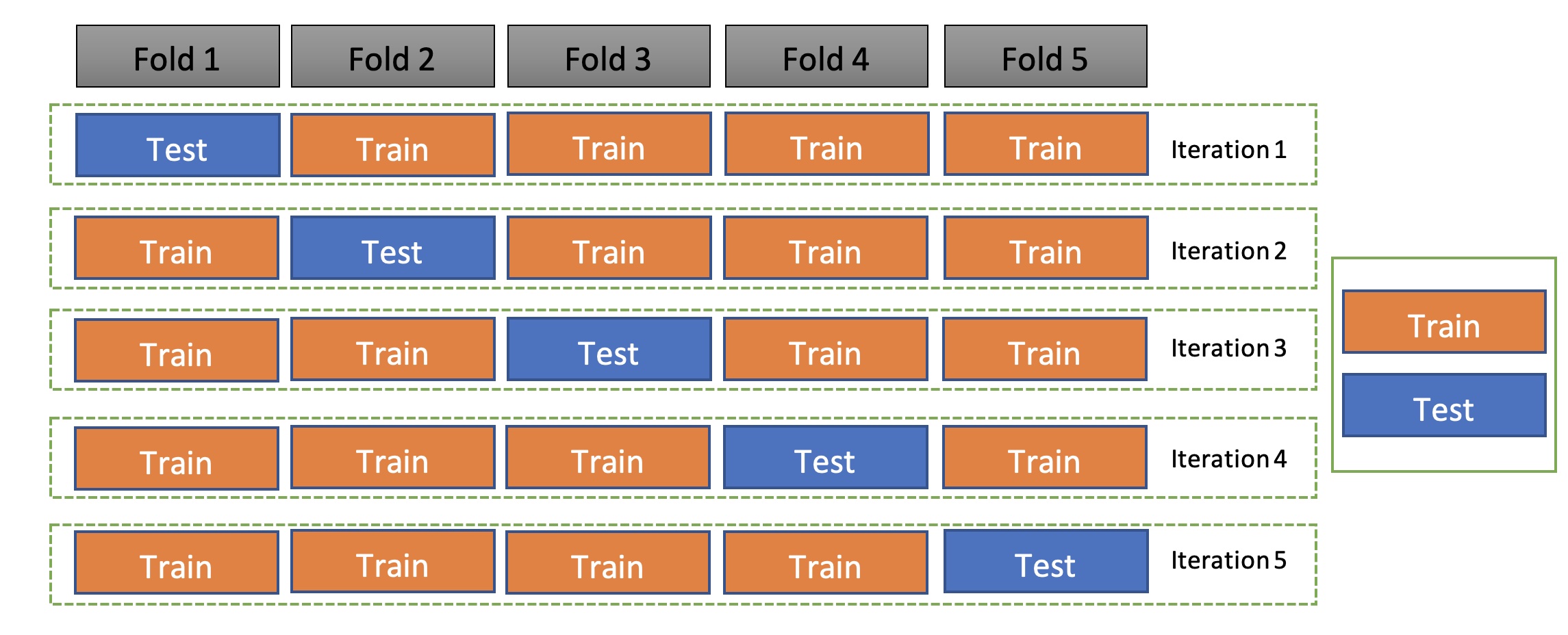}
    \caption{schematic of 5-fold cross validation}
\label{fig:kfold}
\vspace{-.3 cm}
\end{figure}

We use K-fold cross-validation in our bitcoin price prediction model. In K-fold cross-validation, the data is split into K equally sized folds, and the model is trained and evaluated K times, with each fold serving as the validation set once. K-fold cross-validation enables us to gauge the efficiency of our model on data that it has not encountered before, and this is crucial for preventing the model from over-fitting to the training data. By training and evaluating the model on various subsets of the data, we can obtain a more precise assessment of the model's actual performance.
In our specific case, we used 5-fold cross-validation because it strikes a good balance between variance reduction and computational efficiency. Figure~\ref{fig:kfold} shows the schematic of 5-fold cross validation. By using 5 folds, we were able to obtain a reasonable estimate of the model's performance while still keeping the computational requirements manageable. Apart from estimating the performance of the model, K-fold cross-validation is also advantageous in ensuring that our model has the ability to generalize effectively on fresh data. By employing multiple folds and training the model on varying subsets of the data, we can acquire a better understanding of how proficient the model will be in handling new data.

This is crucial as real-world situations may involve data that differs from the data on which the model was trained. Utilizing K-fold cross-validation instills confidence that our model will perform well on novel and unseen data, instead of only on the precise data used for training and testing the model.

\subsection{Model Training and Validation} 
The architecture of the two models used in this project for predicting the future price of Bitcoin: Long Short-Term Memory (LSTM) and Gated Recurrent Unit (GRU). Both LSTM and GRU are types of recurrent neural networks (RNNs) that are designed to handle time series data. They are particularly well-suited for tasks that involve capturing long-term dependencies and patterns in sequential data. In this project, we will use LSTM and GRU to learn the patterns in Bitcoin prices over time and predict future prices based on past prices and other relevant features. Before discussing the specific architecture of each model, we will provide a brief overview of RNNs and their limitations, which motivate the need for LSTM and GRU.

Recurrent neural networks (RNNs) are a type of neural network designed for sequential data processing. Each hidden layer's output in an RNN is fed back as input to the same layer in the next time step, allowing the network to maintain a "memory" of previous inputs and use this information for future predictions.

However, RNNs have limitations that make them difficult to use for certain tasks, such as the "vanishing gradient problem." This problem occurs when gradients become too small during back propagation to capture long-term dependencies, which is more challenging in RNNs due to the multiplication of gradients at each time step. To overcome this limitation, specialized variants of RNNs such as LSTM and GRU have been proposed, which use gating mechanisms to selectively store and retrieve information from the past. This enables them to better handle long-term dependencies and capture complex patterns in sequential data. In the following sections, we will explore the architecture of LSTM and GRU and how they address the limitations of traditional RNNs.

\subsubsection{LSTM}
The LSTM architecture consists of several interconnected layers, including an input layer, an output layer, and a set of "memory cells" that maintain information over time. Each memory cell is controlled by three gates - an input gate, an output gate, and a forget gate - that regulate the flow of information into and out of the cell. These gates are implemented using sigmoid activation functions and element-wise multiplication operations, which allow the network to selectively add or remove information from the memory cells based on the input data. The LSTM architecture can be mathematically represented using a set of equations that describe how the inputs and outputs of each layer are computed. These equations involve matrix multiplications and nonlinear activation functions, and can be trained using backpropagation through time to optimize the network's parameters.
he equations for an LSTM cell are as follows:

Input gate:

$i_t = \sigma(W_{xi}x_t + W_{hi}h_{t-1} + W_{ci}c_{t-1} + b_i)$

Forget gate:

$f_t = \sigma(W_{xf}x_t + W_{hf}h_{t-1} + W_{cf}c_{t-1} + b_f)$

Candidate state:

$\tilde{c}t = \tanh(W{xc}x_t + W_{hc}h_{t-1} + b_c)$

Cell state:

$c_t = f_t * c_{t-1} + i_t * \tilde{c}_t$

Output gate:

$o_t = \sigma(W_{xo}x_t + W_{ho}h_{t-1} + W_{co}c_t + b_o)$

Hidden state:

$h_t = o_t * \tanh(c_t)$

Where:

    $x_t$ is the input at time step $t$, $h_{t-1}$ is the hidden state at time step $t-1$, $c_{t-1}$ is the cell state at time step $t-1$, $i_t$, $f_t$, $o_t$ are the input, forget and output gates respectively, $\tilde{c}_t$ is the candidate state, $W$ and $b$ are the weights and biases of the network, respectively, $\sigma$ is the sigmoid activation function, $*$ denotes element-wise multiplication, and $\tanh$ is the hyperbolic tangent activation function. the summarise of the parameters can be found here

\begin{table}[]
\centering
\begin{tabular}{|l|l|}
\hline
xt   & The input at time step t                   \\ \hline
ht1 & The hidden state at time step t1           \\ \hline
ct1 & The cell state at time step t1             \\ \hline
it   & The input gate                             \\ \hline
ft   & The forget gate                            \\ \hline
ot   & The output gate                            \\ \hline
t    & The candidate cell state                   \\ \hline
W    & The weight matrix                          \\ \hline
b    & The bias vector                            \\ \hline
sigma    & The sigmoid activation function            \\ \hline
*    & Element-wise multiplication                \\ \hline
tanh & The hyperbolic tangent activation function \\ \hline
\end{tabular}
\end{table}

\begin{figure}[h]
    \centering
    \includegraphics[width=.5\textwidth, height=0.3\textwidth]{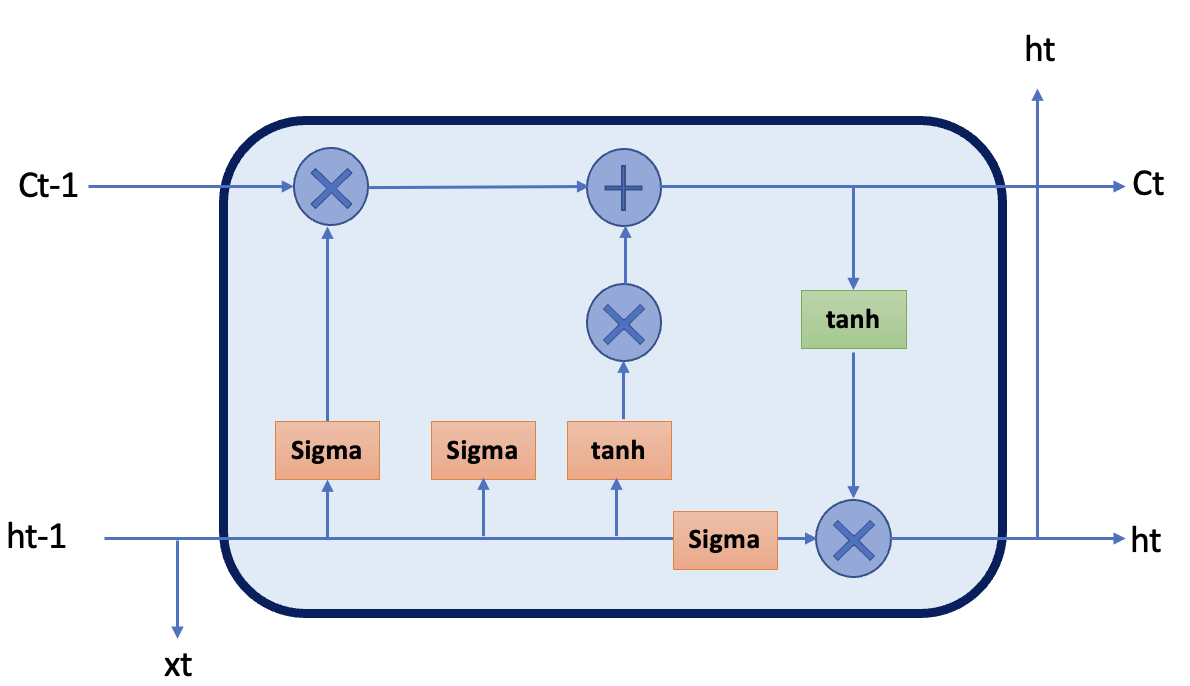}
    \caption{The structure of a long short-term memory (LSTM) cell.}
\end{figure}

\subsubsection{GRU}
GRU are a type of recurrent neural network that were introduced as an alternative to LSTMs. GRUs have a simpler architecture than LSTMs, with fewer parameters, and can be faster to train. Like LSTMs, GRUs also have a gating mechanism that allows them to selectively learn which information to keep or discard.

The architecture of a GRU consists of two gates: an update gate and a reset gate. The update gate determines how much of the past hidden state should be preserved and how much of the new input should be added to the new hidden state. The reset gate determines how much of the past hidden state should be forgotten when computing the new hidden state.

The formulas for the GRU update and reset gates, the candidate activation vector, and the new hidden state are as follows:

Update gate:

$z_t = \sigma(W_z x_t + U_z h_{t-1})$

Reset gate:

$r_t = \sigma(W_r x_t + U_r h_{t-1})$

Candidate activation vector:

$\tilde{h_t} = \tanh(W x_t + U(r_t * h_{t-1}))$

New hidden state:

$h_t = (1 - z_t) * h_{t-1} + z_t * \tilde{h_t}$

Here, $x_t$ is the input at time step $t$, $h_{t-1}$ is the hidden state at time step $t-1$, $z_t$ is the update gate output, $r_t$ is the reset gate output, $\tilde{h_t}$ is the candidate activation vector, $W$ and $U$ are the weights, $\sigma$ is the sigmoid activation function, $*$ denotes element-wise multiplication, and $\tanh$ is the hyperbolic tangent activation function.

\begin{figure}[h]
    \centering
    \includegraphics[width=.5\textwidth, height=0.3\textwidth]{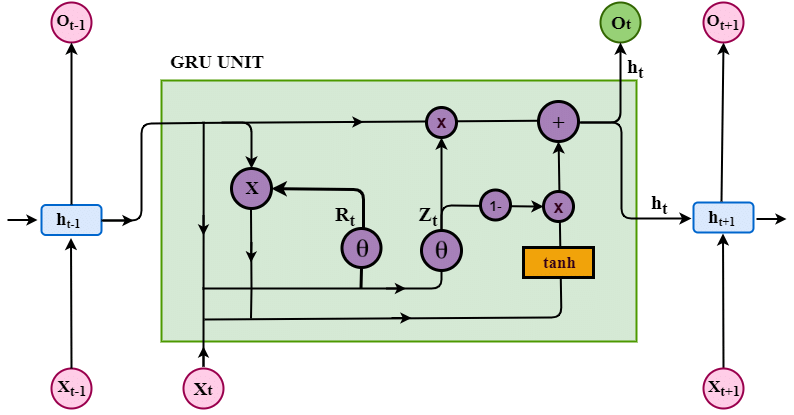}
    \caption{The structure of GRU cell.}
\end{figure}

\subsection{Models Implementation}
To implement the LSTM and GRU models for Bitcoin price prediction, we used Python with the Keras library. Keras is a high-level neural networks API, written in Python and capable of running on top of TensorFlow, CNTK, or Theano.

To ready the data for training and testing, we utilized a 5-fold cross-validation technique. This method entails partitioning the data into five equally-sized folds, utilizing four of the folds as the training set, and leaving one fold for validation in each iteration. We performed this process five times to ensure that each fold is used as the validation set once. This strategy guarantees that the models are assessed on all aspects of the data, and that the outcomes are not skewed toward any particular segment of the data set.

To prevent overfitting and enhance the generalization of models in machine learning, regularization is a popular approach. For our project, we have decided to implement L2 regularization in our neural network models.

The main reason for employing regularization is to lower the level of noise in the data and simplify the model, which can prevent overfitting. Specifically, by using L2 regularization, a penalty term is incorporated into the loss function that promotes small weights in the network. This approach reduces the impact of outliers and other forms of noise in the data, making the model more robust and able to generalize to unseen data.

L2 regularization provides several benefits, including improving the stability of the model. This means that the model becomes less sensitive to small changes in the input data, which is particularly important in financial forecasting where even minor fluctuations in the input data can have a significant impact on the output. The regularization also reduces the complexity of the model, which can lead to better generalization and increased accuracy. Additionally, L2 regularization can help to prevent overfitting, which can occur when the model becomes too complex and begins to fit the noise in the training data rather than the underlying patterns. Overall, L2 regularization is a powerful tool for improving the performance and stability of neural network models.

In order to incorporate L2 regularization into our models, we introduced a penalty term to the loss function that penalizes large weights in the model. This penalty term is directly proportional to the square of the weights, meaning that larger weights will be penalized more heavily than smaller ones. By adjusting the strength of the regularization with a hyperparameter, we were able to fine-tune the models to find the optimal level of regularization that minimized overfitting while still maintaining good performance on the validation and test sets. We also utilized cross-validation to determine the optimal value of the hyperparameter and ensure that the regularization was applied consistently across all folds of the data.

The L2 regularization term can be expressed mathematically as follows:

$L2\ regularization\ term = \lambda\sum_{i=1}^{n}w_i^2$

where $\lambda$ is the regularization parameter and $w_i$ is the weight of the $i^{th}$ feature in the model. The summation is taken over all the features. This term is added to the original loss function of the model, and the strength of the regularization is controlled by the value of $\lambda$.

\section{RESULT AND DISCUSSION} 

Our results showed that the GRU model outperformed the LSTM model in predicting the price of Bitcoin. When we compared the predicted prices with the actual prices in the test set data, the GRU model had a lower MSE than the LSTM model. Specifically, the GRU model had an MSE of 4.67 while the LSTM model had an MSE of 6.25. This indicates that the GRU model was better at capturing the underlying patterns in the data and making accurate predictions. We also generated a plot to visualize the performance of our models. As seen in Fig 6, also we observed that GRU is 30\% faster than LSTM in performance.

\begin{figure}[h]
    \centering
    \includegraphics[width=.5\textwidth, height=0.3\textwidth]{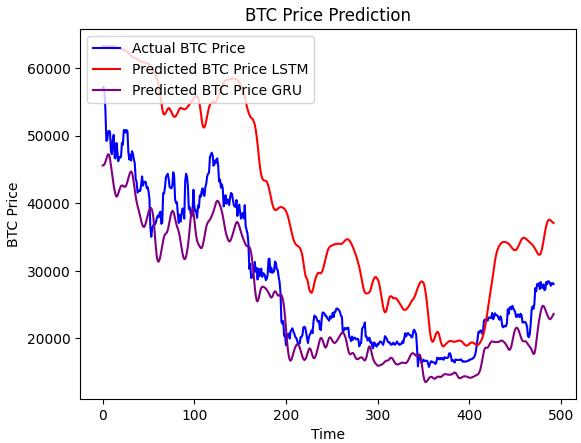}
    \caption{Final Resultant graph of LSTM and GRU that are compered with actual Bitcoin price.}
\end{figure}

The plot of the training and validation MSE loss for our LSTM and GRU models in Fig 7 and 8 showed that both models achieved good performance without overfitting to the training data or producing a lot of noise in the validation data.

Specifically, the plot revealed that the training loss decreased steadily over time for both models, indicating that the models were learning from the data and becoming more accurate at predicting Bitcoin prices. The validation loss also decreased for both models, albeit with some fluctuations, suggesting that the models were generalizing well to new data and were not overfitting.

\begin{figure}[h]
    \centering
    \includegraphics[width=.5\textwidth, height=0.3\textwidth]{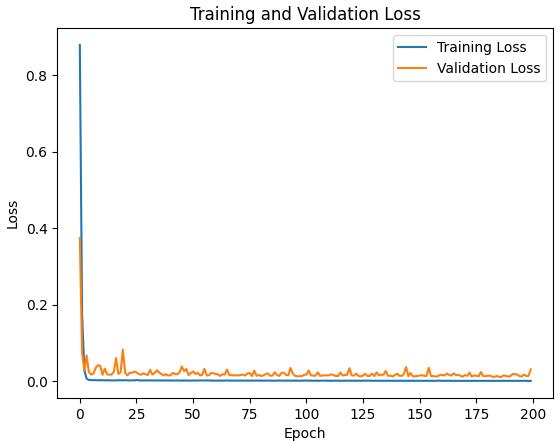}
    \caption{Training and Validation MSE Loss for LSTM.}
\end{figure}

\begin{figure}[h]
    \centering
    \includegraphics[width=.5\textwidth, height=0.3\textwidth]{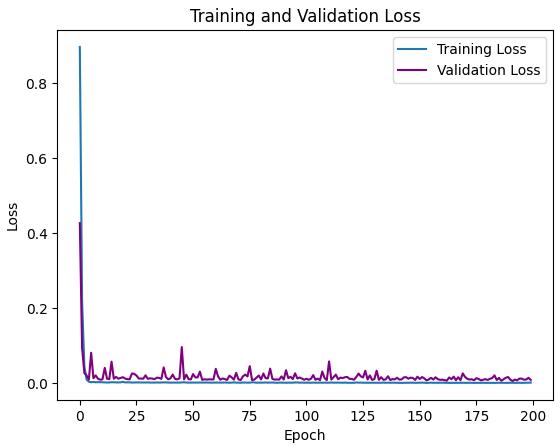}
    \caption{Training and Validation MAE Loss for GRU.}
\end{figure}

\subsection{Discussion}

In our experiments, using the GRU model was more effective than using the LSTM model when predicting Bitcoin price. This aligns with previous research that has highlighted the superiority of GRU models for specific tasks. The explanation for this may be that GRU models are better equipped to process sequential data that has long-term dependencies, as is the case with financial time series data such as Bitcoin prices. Furthermore, our models benefited from L2 regularization which aided in reducing overfitting and noise, ultimately improving performance. As such, L2 regularization can be a valuable tool in optimizing neural network models.

Despite the promising results of our study, it is important to acknowledge several limitations. Firstly, our investigation focused solely on the performance of two neural network models, LSTM and GRU, and there may be other models that could potentially produce superior results. Secondly, our models' performance could have been impacted by the specific parameters we used, such as the learning rate, the number of hidden layers, and the batch size. Further research is required to explore the impact of various parameters on the efficacy of neural network models in predicting Bitcoin prices.

\section{Conclusion and Future Works}
In conclusion, our study highlights the potential of using neural network models for predicting the price of Bitcoin. Our results indicate that the GRU model is more effective than the LSTM model for this task, and that L2 regularization can be a useful technique for improving the performance of neural network models. Our study provides insights into how to improve the accuracy of Bitcoin price predictions, and suggests avenues for further research in this area

\bibliographystyle{IEEEtran}
\bibliography{conference_101719}
\end{document}